\documentclass[12pt]{article}
\usepackage{amsmath}
\usepackage{amssymb}
\usepackage{amsfonts}
\usepackage{epsfig}
\usepackage{mathrsfs}
\usepackage{amsmath,amsthm,amssymb,amscd}
\usepackage{MnSymbol}
\usepackage{rotating}
\usepackage{multirow}
\usepackage{float}
\usepackage[english]{babel}
\usepackage{floatpag}
\usepackage{enumerate}
\usepackage{threeparttable}
\usepackage{color}
\usepackage{tabularx}
\usepackage{booktabs}
\usepackage{lscape}
\usepackage{cite}
\usepackage{algorithm}
\usepackage{algorithmic}
\usepackage{makecell}
\usepackage{tabularray}
\usepackage{graphicx}
\usepackage{tabularray}
\usepackage{hyperref}
\hypersetup{
hidelinks,
colorlinks=true,
linkcolor=blue,
citecolor=blue,
urlcolor = blue
}

\newcommand\appendix@section[1]{\refstepcounter{section}\orig@section*{#1}\addcontentsline{toc}{section}{#1}}
\let\orig@section\section\g@addto@macro\appendix{\let\section\appendix@section}
\makeatother \makeatletter
\renewcommand\footnoterule{\kern-3\p@ \hrule width 1\columnwidth \kern 2.6\p@}
\newskip\@footindent
\renewcommand\@footindent{0pt}
\@addtoreset{footnote}{page}
\long\def\@makefntext#1{\@setpar{\@@par\@tempdima \hsize
\advance\@tempdima-\@footindent \parshape \@ne \@footindent
\@tempdima}\par \noindent \hbox to
\z@{\hss\@thefnmark£º\hspace{0.2em}}#1}

\def\@makefnmark{\hbox{\textsuperscript{\@thefnmark}}}
\makeatother

\newcommand{\sn}{\sum_{i=1}^n}

\newtheorem{thm}{Theorem}
\newtheorem{cor}{Corollary}

\title{Statistical inference for case-control logistic regression via integrating external summary data}
\author{ \vspace{5mm} Hengchao Shi\\
Department of Statistics and Data Science\\
School of Management\\
Fudan University\\
\vspace{1cm}
Shanghai 200433, P.R.China\\
Xinyi Liu\\
School of Mathematical Science\\
Fudan University\\
\vspace{1cm}
Shanghai 200433, P.R.China\\
\and
Ming Zheng \ \ Wen Yu$^\ast$\\
 Department of Statistics and Data Science \\
 School of Management\\
 Fudan University\\
 \vspace{5mm}
 Shanghai 200433, P.R.China\\
 \vspace{1cm} 
{\small $^\ast$Corresponding author: \ wenyu@fudan.edu.cn}}
\date{ }
\begin{document}
\maketitle
\newpage
\begin{abstract}
Case-control sampling is a commonly used retrospective sampling design to alleviate imbalanced structure of binary data. When fitting the logistic regression model with case-control data, although the slope parameter of the model can be consistently estimated, the intercept parameter is not identifiable, and the marginal case proportion is not estimatable, either. We consider the situations in which besides the case-control data from the main study, called internal study, there also exists summary-level information from related external studies. An empirical likelihood based approach is proposed to make inference for the logistic model by incorporating the internal case-control data and external information. We show that the intercept parameter is identifiable with the help of external information, and then all the regression parameters as well as the marginal case proportion can be estimated consistently. The proposed method also accounts for the possible variability in external studies. The resultant estimators are shown to be asymptotically normally distributed. The asymptotic variance-covariance matrix can be consistently estimated by the case-control data. The optimal way to utilized external information is discussed. Simulation studies are conducted to verify the theoretical findings. A real data set is analyzed for illustration.
\end{abstract}

\vfill \hrule \vskip 6pt \noindent {\em MSC:} 62D05 \ 62J12\\
\noindent {\em Some key words}: Case-control sampling; Data integration; Empirical likelihood; Meta-analysis; Retrospective sampling.

\newpage

\section{Introduction}\label{intro}
In many data sets with binary responses, there exists imbalanced data structure in which the subjects belonging to one category, called cases, are rare and the subjects belonging to the other one, called controls, are dominant. The imbalanced structure may bring difficulty for data analysis. One of the possible solutions is to develop retrospective sampling designs for alleviating the imbalanced structure. The case-control design (Mantel \& Haenszel, \textcolor{blue}{\hyperlink{Mantel1959}{1959}}; Miettinen, \textcolor{blue}{\hyperlink{Miettinen1976}{1976}}) is the most commonly used one for such purpose. It takes separate random samples from the case sub-population and control sub-population, respectively. Usually comparable numbers of cases and controls are collected to form a more balanced sample for analysis. It provides a cost-efficient way to alleviate the imbalanced structure, and has unique merit for analyzing binary outcomes.

From model perspective, the logistic regression model is the most fundamental model for fitting binary data. For prospective sampling, the maximum likelihood approach can produce efficient estimation for the regression parameters of the logistic model. Under the case-control sampling, a remarkable finding is that, the score equation derived from the prospective likelihood can still provide a consistent estimator for the slope parameter. However, the intercept parameter can no longer be identified (Farewell, \textcolor{blue}{\hyperlink{Farewell1979}{1979}}; Prentice \& Pyke, \textcolor{blue}{\hyperlink{Prentice1979}{1979}}). This fact implies that one does not need to distinguish the prospective sampling and the case-control sampling if the main interest lies in estimating the covariates' effect, as long as the logistic model is correctly specified for the data. This also contributes one reason why the logistic regression model is treated as the most popular model for fitting binary data.

Though the case-control sampling does not affect the inference on the slope parameter of the logistic model, the intercept parameter can not be identified. Essentially, the case-control design is a response-biased sampling, and the bias is introduced through the intercept parameter. As a result, not only the intercept parameter but also the marginal case percentage is not estimable with a single case-control data. In this era of big data, sometimes several studies for the same research purpose are designed and implemented independently. Thus, when doing analysis based on the collected data of current study, one may have the chance to access information from other similar studies, providing the possibility to improve the analysis of current study. In some literature, the current study conducted by the researchers themselves are called internal study, while other similar studies are called external studies (Zhang et al. \textcolor{blue}{\hyperlink{Zhang2020}{2020}}; Zhang et al. \textcolor{blue}{\hyperlink{Zhang2021}{2021}}; Hu et al. \textcolor{blue}{\hyperlink{Hu2022}{2022}}). The topic of integrating available information from external studies to help the analysis of internal study has received much attention. To fuse summary information on a common parameter from multiple studies, the meta-analysis is the most common method to choose (Singh et al., \textcolor{blue}{\hyperlink{Singh2005}{2005}}; Lin \& Zeng, \textcolor{blue}{\hyperlink{Lin2010}{2010}}; Kundu et al., \textcolor{blue}{\hyperlink{Kundu2010}{2010}}). When the information from external studies is only available in terms of summary statistics, usually this information is utilized by being converted to constraints on the distribution of internal data, c.f., Bickel et al. (\textcolor{blue}{\hyperlink{bickle1993}{1993}}), Qin (\textcolor{blue}{\hyperlink{Qin2000}{2000}}), Qin et al. (\textcolor{blue}{\hyperlink{Qin2015}{2015}}), Chatterjee et al. (\textcolor{blue}{\hyperlink{Chat2016}{2016}}), Zhang et al. (\textcolor{blue}{\hyperlink{Zhang2020}{2020}}), Yuan et al. (\textcolor{blue}{\hyperlink{Yuan2022}{2022}}) etc. Among them, many authors (Qin, \textcolor{blue}{\hyperlink{Qin2000}{2000}}; Qin et al., \textcolor{blue}{\hyperlink{Qin2015}{2015}}; Zhang et al., \textcolor{blue}{\hyperlink{Zhang2020}{2020}}; Yuan et al. \textcolor{blue}{\hyperlink{Yuan2022}{2022}}) used the empirical likelihood proposed by Owen (\textcolor{blue}{\hyperlink{Owen1998}{1998}}, \textcolor{blue}{\hyperlink{Owen1990}{1990}}) to incorporate the constraints, following the idea of Qin \& Lawless (\textcolor{blue}{\hyperlink{Qin1994}{1994}}). Statistical methods for data fusion from external studies with individual-level data available include Yang \& Ding (\textcolor{blue}{\hyperlink{Yang2020}{2020}}), Li et al. (\textcolor{blue}{\hyperlink{Li2021}{2021}}), Sun \& Miao (\textcolor{blue}{\hyperlink{Sun2022}{2022}}), etc. Hu et al. (\textcolor{blue}{\hyperlink{Hu2022}{2022}}) explored the semiparametric efficiency bound for estimation by combining internal study with individual data and external data with summary data. Qin et al. (\textcolor{blue}{\hyperlink{Qin2022}{2022}}) provided a review on statistical methods for data integration from similar studies.

When the internal study contains binary responses sampled by the case-control design, there also exists some research on how to integrate available information from external studies. Qin et al. (\textcolor{blue}{\hyperlink{Qin2015}{2015}}) discussed using auxiliary covariate-specific disease prevalence information to increase the estimation efficiency of the logistic regression for a case-control study. Zhang et al. (\textcolor{blue}{\hyperlink{Zhang2021}{2021}}) developed an integrative analysis of multiple case-control studies, with the individual-level data being available for the internal study and summary data being borrowed from external studies. Tang et al. (\textcolor{blue}{\hyperlink{Tang2021}{2021}}) found that by combining different case-control studies, even for different purposes, can help in improving the estimation efficiency of the slope parameter and identifying the intercept parameter. Quan et al. (\textcolor{blue}{\hyperlink{Quan2024}{2024}}) considered a so-called semi-supervised scenario, where the internal data contains a case-control data and the external data consists of a prospective random sample of the covariates only. They showed that the external data can be integrated with the internal data to identify the intercept parameter and estimate all the parameters efficiently. However, both Tang et al. (\textcolor{blue}{\hyperlink{Tang2021}{2021}}) and Quan et al. (\textcolor{blue}{\hyperlink{Quan2024}{2024}}) assumed that the individual-level data are available for the external studies.

In this paper, we still consider statistical inference for the logistic regression model with case-control internal data, at the same time making use of information from external studies. Our main purpose is to identify the intercept parameter with the help of external information, and then estimate all the model parameters as well as the marginal case proportion. We also explore the possibility to improve the estimation efficiency of the slope parameter. Different from Tang et al. (\textcolor{blue}{\hyperlink{Tang2021}{2021}}) and Quan et al. (\textcolor{blue}{\hyperlink{Quan2024}{2024}}), we do not require that the individual-level data of external studies are available. Instead, we find out that the intercept parameter can still be identified even when only certain summary statistic of the external study is available. In real applications, more often it is much easier to obtain summary data from external studies than individual-level data. To deal with the infinite-dimensional parameter, an empirical likelihood approach following the idea of Qin (\textcolor{blue}{\hyperlink{Qin2000}{2000}}) is adopted. Different from some existing literature (Yuan et al. \textcolor{blue}{\hyperlink{Yuan2022}{2022}}) treating the external information to be at population level, we additionally consider the possible variability in external summary statistics. Specifically, a quadratic term is introduced into the objective function to account for the variability. The estimator for the model parameters is derived from optimizing the proposed objective function and is shown to be consistent and asymptotically normal. With the help of external summary data, one can consistently estimate the intercept parameter of the logistic model as well as the marginal case proportion under the case-control sampling. We also discuss the optimal way to utilize the external information and show that the more external summary information one has, the more efficient estimator one can get for the internal data. Some simulation studies are conducted to show evidence of the theoretical findings.

The rest of the paper is organized as follows. In Section \ref{method}, the proposed methodology is presented. We first give out necessary notation and set up the problem. Then the identifiability issue is discussed and the estimation procedure is described. In Section \ref{thresu}, the large sample properties of the proposed method are derived. The numerical results, including the simulation and real data analysis, are presented in Section \ref{numer}. Section \ref{conclu} concludes. All the technical details are summarized in the Appendix. 

\section{Method}\label{method}

\subsection{Notation and set-up} 

We use $Y$ to denote the binary response variable, taking values $0$ and $1$, and $X$ the $p$-dimensional vector of covariates. The logistic regression model is assumed to relate $Y$ and $X$, that is, 
\begin{eqnarray}\label{logisticModel}
\mathsf{P}(Y=1|X=x)=\frac{\exp\left(\alpha+\beta^\top x\right)}{1+\exp\left(\alpha+\beta^\top x\right)},
\end{eqnarray}
where $\alpha\in\mathbb{R}$ is the intercept parameter and $\beta\in\mathbb{R}^p$ is the slope parameter. We call the sub-population with $Y=1$ the case population and $Y=0$ the control population. Let $F(x)$ be the cumulative distribution function of $X$, and assume that $F(x)$ has a density denoted by $f(x)$. Let $f_1(x)$ and $f_0(x)$ be the conditional density of $X$ given $Y=1$ and $Y=0$, respectively. The internal data with individual-level data is sampled by the case-control design. Specifically, we have a random sample of size $n_1$ from the case population and a random sample of size $n_0$ from the control population. The total size of the internal data is $n=n_1+n_0$. The observed internal data is denoted by $\{(y_i,x_i), i=1\ldots,n\}$.

Besides the internal data, there exists an external data with the size being denoted by $N$. Let $h(x)\in\mathbb{R}^q$ be a known vector function of $x$. For instance, $h(x)$ can be part of $x$ or a transformation of $x$. Suppose that from the external data, we can obtain an estimate of $\mathsf{E}[h(X)]$ as the summary information. For instance, if the external data consists a prospective random sample of the covariates, denoted by $\{x_i, i=n+1,\ldots,n+N\}$, then $\mathsf{E}[h(X)]$ can be consistently estimated by $N^{-1}\sum_{i=n+1}^{n+N}h(x_i)$. We denote the value of $\mathsf{E}[h(X)]$ by $\mu$ and the corresponding estimate by $\tilde{\mu}$, that is, $\tilde{\mu}$ is the available summary data from the external study. We also assume that the external sample size $N$ is known.

\subsection{Identifiability issue}

It is well-known that the intercept parameter $\alpha$ is not identifiable under the case-control sampling. An intuitive argument is given as follows. By using the Bayesian formula, it is easy to see that under the logistic model (\ref{logisticModel}), we have that
\begin{eqnarray}\label{exptilting1}
f_1(x)=\frac{\mathsf{P}(Y=0)}{\mathsf{P}(Y=1)}\exp\left(\alpha+\beta^\top x\right)f_0(x).
\end{eqnarray}
Define $\gamma=\log[\mathsf{P}(Y=0)/\mathsf{P}(Y=1)]$. Then (\ref{exptilting1}) can be written as
\begin{eqnarray}\label{exptilting2}
f_1(x)=\exp\left(\alpha^\ast+\beta^\top x\right)f_0(x),
\end{eqnarray}
where $\alpha^\ast=\gamma+\alpha$. Usually (\ref{exptilting2}) is known as the exponential tilt model (Anderson, 1979; Cheng et al, 2009). Base on the case-control data, both $f_1(x)$ and $f_0(x)$ can be identified as $n_1$ and $n_0$ go to infinity. Thus, as long as $X$ is non-degenerate, $\beta$ and $\alpha^\ast$ can be identified from (\ref{exptilting2}). Note that by this argument, the slope parameter $\beta$ can be identified from the case-control data. However, $\alpha$ cannot be identified because of the existence of $\gamma$. Meanwhile, the case proportion is not identifiable, either. 

When there exists further information on $f(x)$, i.e., the distribution of the covariate, $\alpha$ may become identifiable. Suppose that one can identify the value of $\mu$ from the external study. Let $\mu^1$ and $\mu^0$ denote the conditional expectation of $h(X)$ given $Y=1$ and $Y=0$, respectively, that is, $\mu^1=\int_{\cal X}h(x)f_1(x)dx$ and $\mu^0=\int_{\cal X}h(x)f_0(x)dx$, where ${\cal X}$ is the support of $X$. By the law of total expectation, we have that
\begin{eqnarray}\label{totex}
\mathsf{P}(Y=1)\mu^1+\mathsf{P}(Y=0)\mu^0=\mu.
\end{eqnarray}
It is easy to see that $\mathsf{P}(Y=1)=1/[1+\exp(\gamma)]$ and $\mathsf{P}(Y=0)=\exp(\gamma)/[1+\exp(\gamma)]$. Both $\mu^1$ and $\mu^0$ can be identified from the case-control sample as $n_1$ and $n_0$ go to infinity. Thus, as long as $\mu^1\not=\mu^0$ (i.e., $\beta\not=0$), $\gamma$ can be identified from (\ref{totex}). By the identifiability of $\gamma$, the intercept parameter $\alpha$ and the marginal case proportion can also be identified.

\subsection{Estimation} 

The observed likelihood of the internal case-control data is given by
\begin{eqnarray*}
\prod_{i=1}^n\left[f_1(x_i)\right]^{y_i}\left[f_0(x_i)\right]^{1-y_i}.
\end{eqnarray*}
Based on the exponential tilt model (\ref{exptilting2}), the corresponding log-likelihood is 
\begin{eqnarray}\label{loglik1}
\sum_{i=1}^n\left[y_i(\alpha^\ast+\beta^\top x_i)+\log f_0(x_i)\right].
\end{eqnarray}
To deal with the non-parametric part $f_0(x)$, we adopt the empirical likelihood approach proposed by Owen (1988, 1991). Specifically, let $p_i=f_0(x_i)$ for $i=1,\ldots,n$. Since both $f_0(x)$ and $f_1(x)$ are density functions, we should have the constraints $\sum_{i=1}^np_i=1$ and $\sum_{i=1}^n\exp(\alpha^\ast+\beta^\top x_i)p_i=1$. Moreover, the equation (\ref{totex}) provides another constraint which integrates the summary information from the external study. If $\mu$ were known, the constraint takes the form of
\begin{eqnarray}\label{con1}
\sum_{i=1}^n\frac{\exp(\alpha^\ast+\beta^\top x_i)}{1+\exp(\gamma)}h(x_i)p_i+\frac{\exp(\gamma)}{1+\exp(\gamma)}h(x_i)p_i=\mu.
\end{eqnarray}
Given the estimate $\tilde{\mu}$ obtained from the external study, it is natural to use it to replace $\mu$ in (\ref{con1}). Following the discussion in Zhang et al. (2020), if the sample size of the external study is comparable to that of the internal data, the variability in the estimate $\tilde{\mu}$ should not be ignored. To account for such variability, we treat $\tilde{\mu}$ as a random vector instead of an observed value. In most estimation procedures, the resultant estimator $\tilde{\mu}$ is $\sqrt{N}$-consistent and is asymptotically normally distributed, that is, $\sqrt{N}(\tilde{\mu}-\mu)$ converges in distribution to a normal distribution with mean zero and variance-covariance matrix denoted by $V$. Then the kernel part of the log-likelihood function is given by $-N(\tilde{\mu}-\mu)^\top V^{-1}(\tilde{\mu}-\mu)/2$. Similar to Zhang et al. (2020), we consider adding this part to the log-likelihood (\ref{loglik1}). Let $\theta=(\gamma,\alpha^\ast,\beta^\top,\mu^\top)^\top$ be the collection of all the finite dimensional parameters and ${\bf p}=(p_1,\ldots,p_n)^\top$. We propose to estimate $(\theta,{\bf p})$ by maximizing 
\begin{eqnarray}\label{loglik2}
\sum_{i=1}^n\left[y_i(\alpha^\ast+\beta^\top x_i)+\log p_i\right]-N(\tilde{\mu}-\mu)^\top W^{-1}(\tilde{\mu}-\mu)/2
\end{eqnarray}
over $(\theta,{\bf p})$, subject to
\begin{eqnarray}\label{con2}
\sum_{i=1}^np_i=1, \ \sum_{i=1}^n\exp(\alpha^\ast+\beta^\top x_i)p_i=1, \ \sum_{i=1}^n\frac{\exp(\alpha^\ast+\beta^\top x_i)+\exp(\gamma)}{1+\exp(\gamma)}h(x_i)p_i=\mu,
\end{eqnarray}
with $p_i\geqslant0$, $i=1,\ldots,n$, and $W$ being a given $q\times q$ positive-definite matrix. To maximize (\ref{loglik2}) with the constraints in (\ref{con2}), the Lagrange multipliers is applied. Define
\begin{eqnarray*}
H(x;\theta) = \left(
\begin{array}{c}
\exp(\alpha^\ast+\beta^\top x)-1\vspace{2mm}\\
{\displaystyle\frac{\exp(\alpha^\ast+\beta^\top x)+\exp(\gamma)}{1+\exp(\gamma)}}h(x)-\mu
\end{array}
\right).
\end{eqnarray*}
Then the corresponding objective function with Lagrange multipliers becomes
\begin{eqnarray*}
L_W(\theta, {\bf p}; \lambda, \nu)&=&\sum_{i=1}^n\left[y_i(\alpha^\ast+\beta^\top x_i)+\log p_i\right]-N(\tilde{\mu}-\mu)^\top W^{-1}(\tilde{\mu}-\mu)/2\\
& &-n\lambda\left(\sum_{i=1}^np_i-1\right)-n\nu^\top\sum_{i=1}^np_i H(x_i;\theta),
\end{eqnarray*}
where $\lambda$ and $\nu$ are the Lagrange multipliers. By setting $\partial L_W(\theta, {\bf p}; \lambda, \nu)/\partial p_i=0$, $i=1,\ldots,n$, $\partial L_W(\theta, {\bf p}; \lambda, \nu)/\partial\lambda=0$, and $\partial L_W(\theta, {\bf p}; \lambda, \nu)/\partial\nu=0$, it can be shown that $\lambda=1$ and $p_i=n^{-1}(1+\nu^\top H(x_i;\theta))^{-1}$ under the constraints in (\ref{con2}). By plugging these back to the original optimization problem, maximizing (\ref{loglik2}) with the constraints in (\ref{con2}) is equivalent to maximizing
\begin{eqnarray}\label{prologlik1}
l_W(\theta,\nu)=\sum_{i=1}^n\left[y_i(\alpha^\ast+\beta^\top x_i)-\log\left(1+\nu^\top H(x_i;\theta)\right)\right]-N(\tilde{\mu}-\mu)^\top W^{-1}(\tilde{\mu}-\mu)/2
\end{eqnarray}
with respect to $(\theta,\nu)$. Denote $(\hat{\theta}_W,\hat{\nu}_W)=\mbox{argmax}_{\theta,\nu}l_W(\theta,\nu)$. Let $\hat{\theta}_W=(\hat{\gamma}_W, \hat{\alpha}^\ast_W, \hat{\beta}_W^\top, \hat{\mu}_W^\top)^\top$ and $\hat{\nu}_W$ be the estimator of $\theta$ and $\nu$, respectively. The estimator of $p_i$ can be defined as $\hat{p}_{W,i}=n^{-1}(1+\hat{\nu}_W^\top H(x_i;\hat{\theta}_W))^{-1}$, $i=1,\ldots,n$. Moreover, we can estimate the marginal case proportion based on the estimator of $\gamma$, using the relationship that $\mathsf{P}(Y=1)=1/[1+\exp(\gamma)]$.

If $\tilde{\mu}=N^{-1}\sum_{i=n+1}^{n+N}h(x_i)$, it is easy to see that $V=\mathsf{E}[(h(X)-\mu)^{\otimes2}]$, where $a^{\otimes2}=aa^\top$ for a column vector $a$. Without the individual-level data, $V$ can not be estimated from the external data. However, we can utilize the internal data and the proposed estimator to construct an estimator of $V$, given by
\begin{eqnarray}\label{vest}
\hat{V}=\left(1+\exp(\hat{\gamma}_W)\right)^{-1}\sum_{i=1}^n\left[\left(\exp\left(\hat{\alpha}^\ast_W+\hat{\beta}_Wx_i\right)+\exp\left(\hat{\gamma}_W\right)\right)\hat{p}_{W,i}(h(x_i)-\tilde{\mu})^{\otimes2}\right].
\end{eqnarray}

\section{Theoretical results}\label{thresu}

To obtain the large sample properties of the proposed estimation procedure, some regularity conditions are needed. Let $\theta_0=(\gamma_0,\alpha^\ast_0,\beta_0^\top,\mu_0^\top)^\top$ be the true value of $\theta$.

\noindent{\it Condition 1}. \ $\theta_0$ lies in the interior of a compact parameter space denoted by $\Theta$.

\noindent{\it Condition 2}. \ $\mathsf{E}[H(X;\gamma_0,\alpha^\ast_0,\beta_0,\mu)]= 0 $ has a unique solution at $\mu_0$.

\noindent{\it Condition 3}. \ $\mathsf{E}(\|X\|^3)<\infty$ and $\mathsf{E}[\|h(X)\|^3]<\infty$, where $\|.\|$ stands for the Euclidean norm. $X$ is not concentrated on a hyperplane of dimension smaller than $p$. 

\noindent{\it Condition 4}. \ The external estimator $\tilde{\mu}$ satisfies that $\sqrt{N}(\tilde{\mu}-\mu_0) \to N(0,V)$ in distribution.

\noindent {\it Condition 1} to {\it Condition 3} are quite common regularity conditions. {\it Condition 4} is also reasonable for many commonly used statisitcal anlaysis procedures. Denote the collection of $\hat{\theta}_W$ and $\hat{\nu}_W$ by $\hat{\eta}_W$, that is, $\hat{\eta}_W=(\hat{\theta}_W^\top,\hat{\nu}_W^\top)^\top$. Let $\eta=(\theta^\top,\nu^\top)^\top$ and $\eta_0=(\theta_0^\top,\nu_0^\top)^\top$, where the first dimension of $\nu_0 = n_1/n$ and others equal to 0. The following theorem, proved in Appendix \ref{a1}, gives out the consistency of the proposed estimator $\hat{\theta}_W$.

\begin{thm}\label{thm1}
Suppose that $n/N\to c\in(0,\infty)$ as $n\to\infty$. Under {\it Condition 1} to {\it Condition 4}, with probability one, $\hat{\eta}_W$ satisfies that $\partial l_W(\hat{\eta}_W)/\partial\eta=0$ and $\hat{\theta}_W$ is in the interior of $\{\theta: \|\theta-\theta_0\|\leqslant n^{-1/3}\}$. 
\end{thm}

We then give out the asymptotic distribution of $\hat{\theta}_W$ in the following theorem.

\begin{thm}\label{thm2}
Suppose that $n/N\to c\in(0,\infty)$ as $n\to\infty$. Under {\it Condition 1} to {\it Condition 4}, we have that $\sqrt{n}(\hat{\theta}_W-\theta_0)\to N(0,\Sigma_W)$ in distribution, where the specific form of $\Sigma_W$ is given in Appendix \ref{a1}. 
\end{thm}

\noindent The proof of Theorem \ref{thm1} and \ref{thm2} are given in Appendix \ref{a1}. The asymptotic variance-covariance matrix $\Sigma_W$ can be consistently estimated by the plugged-in method. The specific form of the estimator, denoted by $\hat{\Sigma}_W$, is shown in Appendix \ref{a2}. 

Based on Theorem \ref{thm2}, we can further derive the following optimal property of $\Sigma_W$.

\begin{cor}\label{cor1}
Under {\it Condition 1} to {\it Condition 4}, the variance-covariance matrix of the limiting distribution of $\sqrt{n}(\hat{\theta}_W-\theta_0)$ attains the minimum at $W=V$, with the specific form being given in Appendix A.1. Moreover, the variance-covariance matrix remains the same if $W$ is chosen to be a consistent estimator of $V$.	
\end{cor}

\noindent Corollary \ref{cor1} shows that by setting $W$ to be a consistent estimator of $V$, the estimation efficiency can be improved compared with arbitrary choice of $W$. Note that $\hat{V}$ in (\ref{vest}) is a consistent estimator of $V$. This leads to an iterative algorithm to obtain an estimator achieving the optimal efficiency.

\begin{algorithm}
\caption{Iterated algorithm for achieving optimal estimation efficiency}  
\label{alg1}  
\begin{algorithmic}  
\STATE {Choose a positive-definite matrix $W$, e.g., the $q$-dimensional identity matrix. Solve $(\hat{\theta}_W,\hat{\nu}_W)=\mbox{argmax}_{\theta,\nu}l_W(\theta,\nu)$ and calculate $\hat{p}_{W,i}$, $i=1,\ldots,n$. Calculate $\hat{V}$ at $\hat{\theta}_W$ and $\hat{p}_{W,i}$'s.}   
\REPEAT 
\STATE Step 1. Resolve $(\hat{\theta}_{\hat{V}},\hat{\nu}_{\hat{V}})=\mbox{argmax}_{\theta,\nu}l_{\hat{V}}(\theta,\nu)$ and calculate $\hat{p}_{\hat{V},i}$, $i=1,\ldots,n$.
\STATE Step 2. Update $\hat{V}$ at $\hat{\theta}_{\hat{V}}$ and $\hat{p}_{\hat{V},i}$'s.
\UNTIL{$\hat{\theta}_{\hat{V}}$ is converged.}	
\end{algorithmic} 
\end{algorithm}

\noindent In our numerical studies, we find that usually the algorithm converges quickly in several steps. The asymptotic variance-covariance matrix of $\hat{\theta}_{\hat{V}}$ can be estimated by $\hat{\Sigma}_{\hat{V}}$.

Intuitively, the number of components in $h(x)$, i.e., $q$, also has influence on the estimation efficiency, as illustrated in the following corollary.

\begin{cor}\label{cor2}
The variance-covariance matrix of the limiting distribution of $\sqrt{n}(\hat{\theta}_{\hat{V}}-\theta_0)$ cannot decrease if any function in $h(x)$ is dropped. 	
\end{cor}

\noindent The corollary means that the more external summary information one has, the more efficient estimator one can obtain for the internal data in the asymptotic sense. This is a quite expectable result in the context of empirical likelihood. More external summary information results in larger $q$, so the number of the constraints in (\ref{con2}) increases, making the maximum empirical likelihood estimator to be asymptotically more efficient.





When $n/N \to 0$ as $n\to\infty$, that is, the sample size of the external data is much larger than that of the internal data, the variability of the external data can be neglected. In other words, the summary information $\tilde{\mu}$ can be treated as the population-level parameter. In this case, there is no need to consider the variability in external data and the proposed empirical likelihood approach can be simplified to maximize
\begin{eqnarray*}\label{loglik3}
\sum_{i=1}^n\left[y_i(\alpha^\ast+\beta^\top x_i)+\log p_i\right]
\end{eqnarray*}
over $\gamma$, $\alpha^\ast$, $\beta$, and $p_i$, $i=1,\ldots,n$, subject to
\begin{eqnarray*}
\sum_{i=1}^np_i=1, \ \sum_{i=1}^n\exp(\alpha^\ast+\beta^\top x_i)p_i=1, \ \sum_{i=1}^n\frac{\exp(\alpha^\ast+\beta^\top x_i)+\exp(\gamma)}{1+\exp(\gamma)}h(x_i)p_i=\tilde{\mu},
\end{eqnarray*}
with $p_i\geqslant0$, $i=1,\ldots,n$.  Let $\hat{\gamma}$, $\hat{\alpha}^\ast$, $\hat{\beta}$, and $\hat{p}_i$, $i=1,\ldots,n$, be the resultant maximizer. Denote the collection of the internal parameter estimators by $\hat{\theta}_I=(\hat{\gamma}, \hat{\alpha}^\ast, \hat{\beta}^\top)^\top$ and the true value by $\theta_{I,0}=(\gamma_0,\alpha^\ast_0,\beta_0^\top)^\top$. It can be shown that $\sqrt{n}(\hat{\theta}_I-\theta_{I,0})\to N(0,\Sigma_I)$ in distribution, where the specific form of $\Sigma_I$ is given in Appendix \ref{a1}.

\section{Numerical results}\label{numer}

\subsection{Simulation studies}

Some simulation studies are carried out to assess the finite sample performance of the proposed approach. A 2-dimensional covariates vector $X=(X_1, X_2)$, where both $X_1$ and $X_2$ are independently generated form standard normal distribution, is considered. Given $X$, the response $Y$ is generated from the logistic model (\ref{logisticModel}). We use $\mathsf{p}$ to stand for the marginal case proportion, i.e., $\mathsf{P}(Y=1)$. The case-control sampling is used to form the internal data, with $n_1$ cases and $n_0$ controls. We use $\mathsf{q}$ to stand for the case proportion, i.e., $\mathsf{q}=n_1/n$. The external data consists of $N$ independent and identically distributed samples from the distribution of $X$. 

We consider different choice of the regression parameters $\alpha$, $\beta_1$, and $\beta_2$, resulting in different values of $\mathsf{p}$. Also, different values of $\mathsf{q}$, $n$, and $N$ are tried. Six random number generating schemes with different parameter values we consider are listed in Table \ref{para_set}. We set $h(X)=X$, so the available external information $\tilde{\mu}$ is just the sample means of the covariates in the external data.  With the external data information, we use the proposed method to obtain the maximal empirical likelihood estimator of the regression parameters and the estimated standard errors. Two kinds of estimators are calculated. The first one is $\hat{\theta}_W$ with $W$ being a diagonal matrix with diagonal elements $0.2$ and $2$, which means that $W$ is an arbitrarily selected matrix to be put in (\ref{loglik2}). The second one is $\hat{\theta}_{\hat{V}}$ proposed in Algorithm \ref{alg1}, which has the optimal estimation efficiency. We also consider the situation with no external data and use the internal data only to estimate the regression parameters, that is, to fit model (\ref{logisticModel}) by a single case-control data. The random numbers are generated for 1000 replications. For each regression parameter's estimator, we record the average bias, the empirical standard error, and average of the estimated standard errors, and the empirical coverage percentage of the 95\% Wald confidence interval. We also record the average bias of the estimator of the marginal case proportion. The results are listed in Table \ref{simul_est}.

\begin{table}
\centering
\caption{Parameter values for random data generating scheme}
\bigskip
\begin{tabular}{cccccccc}
\toprule
Scheme & $\alpha$ & $\beta_1$ & $\beta_2$ & $n_0$ & $n_1$ & $\mathsf{p}$ & $\mathsf{q}$  \\ \midrule
$A_1$ & $-5$ & $-2$ & 2 & 4000 & 800 & 0.067 & 0.167  \\ 
$A_2$ & $-4$ & 2 & 1 & 4000 & 800 & 0.116 & 0.167  \\ 
$B_1$ & $-5$ & $-2$ & 2 & 3000 & 1500 & 0.067 & 0.333  \\ 
$B_2$ & $-4$ & 2 & 1 & 3000 & 1500 & 0.116 & 0.333  \\ 
$C_1$ & $-5$ & $-2$ & 2 & 2000 & 2000 & 0.067 & 0.500  \\ 
$C_2$ & $-4$ & 2 & 1 & 2000 & 2000 & 0.116 & 0.500 \\
\bottomrule
\end{tabular}
\label{para_set}
\end{table}

\begin{sidewaystable}[]
\centering
\caption{Simulation results for estimation of regression parameters and case proportion}
\bigskip
\resizebox{\columnwidth}{!}{%
\begin{tabular}{cccccccccccccccc}
\toprule
\multirow{2}{*}{Scheme} & \multirow{2}{*}{External data size} & \multirow{2}{*}{Estimator} & \multicolumn{3}{c}{Bias} & \multicolumn{3}{c}{SE} & \multicolumn{3}{c}{ESE} & \multicolumn{3}{c}{CP} & \multirow{2}{*}{$\mathsf{p}$ bias} \\
&     &            & $\alpha$  & $\beta_1$  & $\beta_2$  & $\alpha$  & $\beta_1$ & $\beta_2$ & $\alpha$  & $\beta_1$  & $\beta_2$ & $\alpha$  & $\beta_1$ & $\beta_2$ &         \\
\midrule
\multirow{5}{*}{$A_1$}     & $N=0$ & $\hat{\theta}_{MLE}$           & 1.005  & $-0.005$ & 0.003  & 0.122  & 0.080 & 0.081 & 0.125  & 0.080  & 0.080 & 0.000  & 0.946 & 0.943 & 0.100   \\
& $N=n$ & $\hat{\theta}_W$             & $-0.017$ & $-0.003$ & 0.000  & 0.128  & 0.081 & 0.081 & 0.241  & 0.078  & 0.078 & 0.997  & 0.928 & 0.948 & 0.000   \\
&     & $\hat{\theta}_{\hat{V}}$     & $-0.017$ & $-0.001$ & 0.004  & 0.129  & 0.079 & 0.080 & 0.229  & 0.078  & 0.078 & 0.939  & 0.942 & 0.997 & 0.000   \\
& $N=4n$ & $\hat{\theta}_W$             & $-0.024$ & $-0.006$ & 0.009  & 0.128  & 0.082 & 0.081 & 0.196  & 0.078  & 0.078 & 0.993  & 0.929 & 0.938 & 0.000   \\
&     & $\hat{\theta}_{\hat{V}}$     & $-0.016$ & $-0.004$ & 0.006  & 0.127  & 0.080 & 0.077 & 0.195  & 0.078  & 0.078 & 0.993  & 0.941 & 0.951 & 0.000   \\
\midrule
\multirow{5}{*}{$A_2$}     & N=0 & $\hat{\theta}_{MLE}$           & 0.420  & 0.003  & 0.006  & 0.110  & 0.084 & 0.083 & 0.110  & 0.080  & 0.080 & 0.049  & 0.943 & 0.948 & 0.051   \\
& $N=n$ & $\hat{\theta}_W$             & 0.002  & 0.000  & 0.002  & 0.109  & 0.078 & 0.076 & 0.176  & 0.078  & 0.078 & 0.995  & 0.951 & 0.953 & 0.000   \\
&     & $\hat{\theta}_{\hat{V}}$     & $-0.006$ & 0.005  & 0.007  & 0.118  & 0.080 & 0.081 & 0.167  & 0.078  & 0.078 & 0.948  & 0.938 & 0.984 & 0.000   \\
& $N=4n$ & $\hat{\theta}_W$             & $-0.005$ & 0.005  & 0.007  & 0.114  & 0.078 & 0.079 & 0.148  & 0.078  & 0.078 & 0.984  & 0.948 & 0.946 & 0.000   \\
&     & $\hat{\theta}_{\hat{V}}$     & 0.002  & 0.001  & 0.005  & 0.116  & 0.080 & 0.082 & 0.147  & 0.078  & 0.078 & 0.981  & 0.939 & 0.940 & 0.000   \\
\midrule
\multirow{5}{*}{$B_1$}     & $N=0$ & $\hat{\theta}_{MLE}$           & 1.920  & $-0.006$ & 0.008  & 0.096  & 0.070 & 0.070 & 0.101  & 0.071  & 0.071 & 0.000  & 0.955 & 0.951 & 0.267   \\
& $N=n$ & $\hat{\theta}_W$             & $-0.020$ & $-0.006$ & 0.004  & 0.111  & 0.071 & 0.072 & 0.241  & 0.069  & 0.069 & 1.000  & 0.942 & 0.935 & 0.000   \\
&     & $\hat{\theta}_{\hat{V}}$     & $-0.018$ & $-0.005$ & 0.004  & 0.106  & 0.072 & 0.071 & 0.228  & 0.068  & 0.068 & 0.942  & 0.940 & 0.998 & 0.000   \\
& $N=4n$ & $\hat{\theta}_W$             & $-0.020$ & $-0.003$ & 0.008  & 0.109  & 0.072 & 0.069 & 0.193  & 0.068  & 0.068 & 0.993  & 0.930 & 0.950 & 0.000   \\
&     & $\hat{\theta}_{\hat{V}}$     & $-0.016$ & $-0.002$ & 0.005  & 0.109  & 0.074 & 0.072 & 0.189  & 0.068  & 0.068 & 0.995  & 0.923 & 0.943 & 0.000   \\
\midrule
\multirow{5}{*}{$B_2$}     & $N=0$ & $\hat{\theta}_{MLE}$           & 1.338  & 0.004  & 0.004  & 0.083  & 0.070 & 0.072 & 0.088  & 0.070  & 0.070 & 0.000  & 0.956 & 0.947 & 0.217   \\
& $N=n$ & $\hat{\theta}_W$             & 0.005  & 0.004  & 0.001  & 0.096  & 0.072 & 0.069 & 0.168  & 0.069  & 0.069 & 0.993  & 0.944 & 0.940 & 0.000   \\
&     & $\hat{\theta}_{\hat{V}}$     & $-0.002$ & 0.004  & 0.005  & 0.098  & 0.072 & 0.069 & 0.158  & 0.069  & 0.069 & 0.947  & 0.950 & 0.997 & 0.000   \\
& $N=4n$ & $\hat{\theta}_W$             & 0.001  & 0.004  & 0.003  & 0.097  & 0.071 & 0.073 & 0.136  & 0.068  & 0.068 & 0.984  & 0.940 & 0.935 & 0.000   \\
&     & $\hat{\theta}_{\hat{V}}$     & $-0.001$ & 0.003  & 0.003  & 0.096  & 0.071 & 0.072 & 0.136  & 0.068  & 0.068 & 0.981  & 0.939 & 0.936 & 0.000   \\
\midrule
\multirow{5}{*}{$C_1$}     & $N=0$ & $\hat{\theta}_{MLE}$           & 2.625  & $-0.001$ & $-0.002$ & 0.092  & 0.074 & 0.072 & 0.094  & 0.072  & 0.072 & 0.000  & 0.949 & 0.956 & 0.433   \\
& $N=n$ & $\hat{\theta}_W$             & $-0.017$ & $-0.004$ & 0.005  & 0.101  & 0.071 & 0.071 & 0.267  & 0.069  & 0.069 & 0.999  & 0.930 & 0.950 & 0.000   \\
&     & $\hat{\theta}_{\hat{V}}$     & $-0.015$ & $-0.006$ & 0.003  & 0.101  & 0.071 & 0.073 & 0.254  & 0.069  & 0.069 & 0.943  & 0.935 & 0.998 & 0.000   \\
& $N=4n$ & $\hat{\theta}_W$             & $-0.013$ & $-0.002$ & $-0.001$ & 0.103  & 0.073 & 0.071 & 0.222  & 0.069  & 0.069 & 0.996  & 0.941 & 0.941 & 0.000   \\
&     & $\hat{\theta}_{\hat{V}}$     & $-0.015$ & $-0.006$ & 0.002  & 0.097  & 0.071 & 0.070 & 0.219  & 0.069  & 0.069 & 0.999  & 0.942 & 0.942 & 0.000   \\ \midrule
\multirow{5}{*}{$C_2$}     & $N=0$ & $\hat{\theta}_{MLE}$           & 2.030  & 0.006  & 0.005  & 0.077  & 0.073 & 0.074 & 0.082  & 0.072  & 0.072 & 0.000  & 0.942 & 0.947 & 0.384   \\
& $N=n$ & $\hat{\theta}_W$             & 0.000  & 0.003  & 0.004  & 0.092  & 0.071 & 0.073 & 0.180  & 0.069  & 0.069 & 0.994  & 0.946 & 0.942 & 0.000   \\
&     & $\hat{\theta}_{\hat{V}}$     & 0.002  & 0.002  & 0.004  & 0.093  & 0.072 & 0.072 & 0.171  & 0.069  & 0.069 & 0.938  & 0.947 & 0.994 & 0.000   \\
& $N=4n$ & $\hat{\theta}_W$          & 0.002  & 0.001  & 0.002  & 0.092  & 0.071 & 0.070 & 0.149  & 0.069  & 0.069 & 0.987  & 0.937 & 0.946 & 0.000   \\
&     & $\hat{\theta}_{\hat{V}}$     & 0.004  & 0.000  & 0.000  & 0.090  & 0.070 & 0.071 & 0.148  & 0.069  & 0.069 & 0.989  & 0.942 & 0.945 & 0.000 \\
\bottomrule
\end{tabular}
\label{simul_est}
}
{\footnotesize\begin{tablenotes}
\item[1]Bias: average bias of the estimates; SE: empirical standard error of the estimates; ESE: average of the estimated standard errors; CP: empirical coverage probabilities of Wald-type confidence intervals with $95\%$ confidence level; $\mathsf{p}$ Bias: average bias of the estimator of $\mathsf{P}(Y=1)$; $\hat{\theta}_{MLE}$: Maximum likelihood estimator for single case-control sample; $\hat{\theta}_{W}$: Maximum empirical likelihood estimator with arbitrarily chosen $W$; $\hat{\theta}_{\hat{V}}$: Maximum empirical likelihood estimator defined in Algorithm \ref{alg1}.
\end{tablenotes}}
\end{sidewaystable}

Note that $N=0$ corresponds to the scenario with no external data. From the simulation results, we first see that the case-control sample cannot identify the intercept parameter and marginal case proportion. Without the external information, the bias of estimation in $\alpha$ and $\mathsf{P}(Y=1)$ is obvious. By incorporating external summary information, our method significantly reduces the bias and the proposed empirical likelihood estimator for $\alpha$ is essentially unbiased. Meanwhile, the marginal case proportion can be also consistently estimated. Secondly, we find that the external information not only helps in identifying the intercept parameter, but also slightly improves the estimation efficiency of the slope parameters in some cases. Thirdly, using consistent estimator of $V$ as the choice of $W$ is useful for improving the estimation efficiency of regression parameters, which coincides with our theoretical results. In our simulation studies, we see that the improvement is especially significant in $\alpha$ estimation. Finally, the estimated standard errors are generally quite close to the corresponding empirical standard errors. The Wald confidence intervals also give out satisfactory coverage probabilities compared with the nominal level.

\subsection{A real data example}

We consider a real data-set from the National Institute of Diabetes and Digestive and Kidney disease, called the Pima Indians diabetes data. The response of this data is the variable called ``Outcome" which indicates the presence (coded as 1) and absence (coded as 0) of the diabetes. The other variables in the data can be used as covariates. We use three variables with complete observations, ``Glucose", ``Pregnancies", and ``BMI", as the covariates, and all the covariates are standardized. There are 264 cases and 488 controls in the data-set. The total 752 subjects with four variables are treated as the full data. We randomly split the full data into two parts with equal size. One is used as the internal data and the other as the external data. For the internal data, we randomly sample 100 cases and 100 controls to form a case-control internal data. A logistic regression model is fitted based on the case-control internal data. The external information is the sample means of three covariates, calculated from the external data. We calculate the proposed estimator $\hat{\theta}_{\hat{V}}$ to obtain the regression parameter estimates and the marginal case proportion estimate. Meanwhile, we also obtain the estimation results based on the internal case-control data only. The splitting and sampling procedure are replicated for 100 times. The average of the parameter estimates and estimated standard errors are reported. The estimation results obtained from the full data logistic regression model are treated as the benchmark. All the results are summarized in Table \ref{real_est}.

From the results, we see that the estimate of the intercept parameter from the case-control sample is quite far from that of the full data. By contrast, the estimate incorporating the external information has a much closer value to the full data estimate. By using the proposed approach, the estimate of the marginal case proportion is also close to the full data case percentage. Again, the analysis results indicate our proposed method is useful for eliminating the estimation bias in the intercept parameter of logistic model and the marginal case proportion, meanwhile improving the estimation efficiency of the slope parameters.

\begin{table}[htbp]
\centering
\begin{threeparttable}
\caption{Estimation results for Pima India diabetes data}
\bigskip
\begin{tabular}{ccccccccc}
\toprule
    Method    &       & $\alpha$  & $\beta_1$ & $\beta_2$ & $\beta_3$ & Case proportion  \\ \midrule
\multirow{2}{*}{Internal CC $\hat{\theta}_{MLE}$} & EST.A & 0.052  & 1.211 & 0.471 & 0.690 & \multirow{2}{*}{0.500}   \\
           & ESE.A & 0.176  & 0.218 & 0.184 & 0.203 &  \\[1em]
\multirow{2}{*}{MELE $\hat{\theta}_{\hat{V}}$}  & EST.A & $-0.750$ & 1.059 & 0.496 & 0.577 & \multirow{2}{*}{0.354}  \\
           & ESE.A & 0.341  & 0.197 & 0.176 & 0.168 &  \\[1em]
\multirow{2}{*}{Full Data $\hat{\theta}_{MLE}$}  & EST   & $-0.835$ & 1.135 & 0.443 & 0.623 & \multirow{2}{*}{0.351}  \\
           & ESE   & 0.096  & 0.106 & 0.092 & 0.101 & \\
\bottomrule
\end{tabular}
\label{real_est}
{\begin{tablenotes}[para,flushleft]
\footnotesize
\item Internal CC $\hat{\theta}_{MLE}$: the maximum likelihood estimator based on the internal case-control data; MELE $\hat{\theta}_{\hat{V}}$: the proposed maximum empirical likelihood estimator; Full Data $\hat{\theta}_{MLE}$: the maximum likelihood estimator based on the full data;  EST.A: average of estimated parameter values; ESE.A: average of estimated standard errors; EST: estimated parameter value; ESE: estimated standard error.
\end{tablenotes}}
\end{threeparttable}
\end{table}

\section{Concluding remarks}\label{conclu}

We propose a statistical inference procedure for case-control logistic regression model when there exist external data information. Our approach uses external summary data to identify all the regression parameters when the individual-level internal data is collected by the case-control sampling, a sampling design commonly used for binary response data. For case-control data, the intercept and the marginal case proportion are not identifiable. By using empirical likelihood approach, we incorporate the summary-level external data information with the internal case-control data to identify the intercept parameter as well as the marginal case proportion. In many real applications, the variability of external data cannot be ignored since the sample size of the external data is comparable with that of the internal data. A quadratic form is introduced into the objective function to account for the variability. The consistency and asymptotic normality of the proposed estimator are established. We also consider improving the estimation efficiency by proper construction of the quadratic form. An algorithm to obtain the more efficient estimator is provided.

The external data integration in statistical literature has a close relationship with transfer learning in machine learning context. The idea of the proposed approach can be extended to some other scenarios, such as the situations where the internal data are sampled by other biased sampling designs, or the responses are no longer binary data. In machine learning, usually the logistic regression models are not correctly specified for the data and used as working models. It is also of interest whether the external information may help in increasing the prediction accuracy under mis-specified logistic models. All these will be the possible topics of our future research.

\appendix

\section{Appendix}

\subsection{Proof of theorems and corollaries}\label{a1}

Here we give out the proof of the theorems and corollaries in Section 3.

\noindent{\it Proof of Theorem \ref{thm1}:} \ It is not difficult to see that optimizing (\ref{prologlik1}) is equivalent to optimize $l_W(\theta) = l_W(\theta, \nu(\theta))$ with respect to $\theta$, where $\nu(\theta)$ is the solution of 
\begin{eqnarray*}
\sn \frac{H(x_i;\theta)}{1+\nu^\top H(x_i;\theta)}=0
\end{eqnarray*}
for $\nu$ (see Qin \& Lawless, 1994). Define $\Theta_n=\{\theta : \left.\left\| \theta-\theta_0 \right\| \leqslant n^{-1 / 3}\right\} \subset \Theta$ for some large $n$. We need to show that the maximal point of $l_W(\theta)$ is attained at an interior point of $\Theta_n$. 

Similar to the proof in Zhang et al. (2020), we first give an upper bound when $\theta$ is on $\partial \Theta_n$, the surface of  $\Theta_n$. For any $\theta \in \partial \Theta_n$, denote a unit vector $u = (\gamma, \alpha^\ast, \beta^\top, \mu^\top)^\top$ satisfying $\theta = \theta_0 + u n^{-1/3}$. We separate the right side of (\ref{prologlik1}) into two parts. Firstly, we consider the following part.
\begin{eqnarray*}
& &\left[\sum_{i=1}^n y_i\left(\alpha^\ast+\beta^{\top} x_i\right)-\frac{N}{2}(\tilde{\mu}-\mu)^{\top} W^{-1}(\tilde{\mu}-\mu)\right]-\left[\sum_{i=1}^n y_i\left(\alpha_0^\ast+\beta_0^{\top} x_i\right)-\frac{N}{2}(\tilde{\mu}-\mu_0)^{\top} W^{-1}(\tilde{\mu}-\mu_0)\right] \\
&=& (\alpha^\ast - \alpha_0)\sum_{i=1}^n y_i + (\beta - \beta_0)^T \sum_{i=1}^n y_i x_i + N(\mu-\mu_0)^{\top} W^{-1}(\tilde{\mu}-\mu_0) - \frac{N}{2}(\mu-\mu_0)^{\top} W^{-1}(\mu-\mu_0)\\
&\leqslant&\left\lVert \alpha^\ast - \alpha_0 \right\rVert \left\lVert \sum_{i=1}^n y_i \right\rVert + \left\lVert \beta - \beta_0 \right\rVert \left\lVert \sum_{i=1}^n y_i x_i \right\rVert + \left\lVert \mu-\mu_0 \right\rVert \left\lVert NW^{-1}(\tilde{\mu}-\mu_0)\right\rVert - \frac{N}{2}(\mu-\mu_0)^{\top} W^{-1}(\mu-\mu_0) \\
&=&O_p (n^{-1/3} \left\lVert \alpha_u^\ast \right\rVert n^{1/2}(\log \log n)^{1/2}) + O_p (n^{-1/3} \left\lVert \beta_u \right\rVert n^{1/2}(\log \log n)^{1/2})+O_p (n^{-1/3}\left\lVert \mu_u \right\rVert  n^{1/2})\\
& &-O_p (n^{1/3}\left\lVert \mu_u \right\rVert^2)\\
&=&O_p(n^{1/3})
\end{eqnarray*}
The equations hold by law of the iterated logarithm theorem (see Theorem A in Section 1.10 of Serfling, 1980), in which we have the fact that $\left\lVert \sum_{i=1}^n y_i \right\rVert = O_p (n^{1/2}(\log \log n)^{1/2}) $ and $\left\lVert \sum_{i=1}^n y_i x_i \right\rVert = O_p (n^{1/2}(\log \log n)^{1/2})$. 

Secondly, for $\nu = \nu(\theta)$ with $ \theta \in \partial \Theta_n$ satisfying $\sum_{i=1}^n H(x_i ; \theta)(1+\nu^{\top}H(x_i ; \theta))^{-1} = 0$, similar to the proof in Owen(1990), we have that
\begin{eqnarray*}
\nu(\theta) &=& \left(\frac{1}{n}\sum_{i=1}^n H(x_i ; \theta) H^\top(x_i ; \theta) \right)^{-1} \left(\frac{1}{n} \sum_{i=1}^n H(x_i ; \theta) \right) + o_p(n^{-1/3})\\
&=& \left[E\{H(X ; \theta_0) H^\top(X ; \theta_0)\}\right]^{-1} \left[\frac{1}{n} \sum_{i=1}^n H(x_i ; \theta_0) + \frac{1}{n} \sum_{i=1}^n \frac{\partial H(x_i ; \theta_0)}{\partial \theta}(\theta - \theta_0)\right] + o_p(n^{-1/3})\\
&=& O_p (n^{-1/2}(\log \log n)^{1/2} + n^{-1/3}) + o_p(n^{-1/3})\\
&=& O_p(n^{-1/3}).
\end{eqnarray*}
Therefore, we have that
\begin{eqnarray*}
& & -\log \{1+\nu^{\top}(\theta) H(x_i ; \theta) \} \\
&=& - \sum_{i=1}^n \nu^{\top}(\theta) H(x_i ; \theta) + \frac{1}{2} \sum_{i=1}^n \{ \nu^{\top}(\theta) H(x_i ; \theta) \}^2 + o_p(n^{1/3})\\
&=& -\frac{n}{2}\left(\frac{1}{n} \sum_{i=1}^n H(x_i ; \theta) \right)^\top \left(\frac{1}{n}\sum_{i=1}^n H(x_i ; \theta) H^\top(x_i ; \theta) \right)^{-1} \left(\frac{1}{n} \sum_{i=1}^n H(x_i ; \theta) \right) + o_p(n^{1/3})\\
&=& -\frac{n}{2} \left[\frac{1}{n} \sum_{i=1}^n H(x_i ; \theta_0) + \frac{1}{n} \sum_{i=1}^n \frac{\partial H(x_i ; \theta_0)}{\partial \theta}(\theta - \theta_0)\right]^\top \left(\frac{1}{n}\sum_{i=1}^n H(x_i ; \theta_0) H^T(x_i ; \theta_0)  \right)^{-1} \\
& &\times \left[\frac{1}{n} \sum_{i=1}^n H(x_i ; \theta_0) + \frac{1}{n} \sum_{i=1}^n \frac{\partial H(x_i ; \theta_0)}{\partial \theta}(\theta - \theta_0)\right] + o_p(n^{1/3}) \\
&=& -\frac{n}{2} \left[O_p (n^{-1/2}(\log \log n)^{1/2}) + \mathsf{E} \left \{ \frac{\partial H(X ; \theta_0)}{\partial \theta} \right\} u n^{-1/3}\right]^\top \left[\mathsf{E}\{H(X ; \theta_0) H^T(X ; \theta_0)\}\right]^{-1} \\
& & \times \left[O_p (n^{-1/2}(\log \log n)^{1/2}) + \mathsf{E} \left \{ \frac{\partial H(X ; \theta_0)}{\partial \theta} \right\} u n^{-1/3}\right] + o_p(n^{1/3}) \\
&\leqslant& -c_1 n^{1/3} + o_p(n^{1/3}) \\
&\leqslant& -c_1 n^{1/3}/2,
\end{eqnarray*}
where $c_1$ is the smallest eigenvalue of 
\begin{eqnarray*}
\mathsf{E}\left \{ \frac{\partial H(X ; \theta_0)}{\partial \theta} \right\}^{\top} \left[\mathsf{E}\{H(X ; \theta_0) H^\top(X ; \theta_0)\}\right]^{-1} \mathsf{E}\left \{ \frac{\partial H(X ; \theta_0)}{\partial \theta} \right\}.
\end{eqnarray*}
Similarly, we have that
\begin{eqnarray*}
& & -\log \{1+\nu^{\top}(\theta_0) H(x_i ; \theta_0) \}\\
&=& -\frac{n}{2}\left(\frac{1}{n} \sum_{i=1}^n H(x_i ; \theta_0) \right)^\top\left(\frac{1}{n}\sum_{i=1}^n H(x_i ; \theta_0) H^{\top}(x_i ; \theta_0) \right)^{-1} \left(\frac{1}{n} \sum_{i=1}^n H(x_i ; \theta_0) \right) + o_p(1)\\
&=& O_p(\log \log n).
\end{eqnarray*}

Finally, combining all the results above, we have that for any $\theta \in \partial \Theta_n$, $l_W(\theta) \leqslant  l_W(\theta_0)$ almost surely. It means that $l_W(\theta)$ attains its maximum value at $\hat{\theta}_W$ in the interior of the ball $\{\theta : \left.\left\|\theta-\theta_0\right\| \leq n^{-1 / 3} \right\}$, and $\hat{\nu}_W$ satisfying $\sum_{i=1}^n H(x_i ; \hat{\theta}_W)(1+\hat{\nu}_W^{\top}H(x_i ; \hat{\theta}_W))^{-1} = 0$. For $\hat{\eta}_W$, it's easy to check that $\partial l_W(\hat{\eta}_W) / \partial \theta = 0$ and $\partial l_W(\hat{\eta}_W) / \partial \nu = 0$, which completes the proof.\qed

\medskip

\noindent{\it Proof of Theorem \ref{thm2}:} \ We define $\delta_i = \exp(\alpha^\ast + \beta^\top x_i), \rho = n_1/n_0, \Delta_i = 1+ \rho \delta_i$.  After some calculation, we get the first order derivation of $l_W(\eta)$  at the true value
\begin{eqnarray*}
\frac{\partial l_W(\eta_0)}{\partial \gamma}=0, \ \frac{\partial l_W(\eta_0)}{\partial \alpha^\ast}=n_1 - \rho \sum_{i=1}^n \frac{\delta_i}{\Delta_i}, \ \frac{\partial l_W(\eta_0)}{\partial \beta}=\sum_{i=1}^n y_i x_i - \rho \sum_{i=1}^n \frac{\delta_i x_i}{\Delta_i}
\end{eqnarray*}
\begin{eqnarray*}
\frac{\partial l_W(\eta_0)}{\partial \mu}= NW^{-1}(\tilde{\mu}-\mu_0), \ \frac{\partial l_W(\eta_0)}{\partial \nu}= -(1+\rho)\sum_{i=1}^n \frac{H(x_i ; \theta_0)}{\Delta_i}.
\end{eqnarray*}
Similarly, the second order derivation at the true value are given as follows.
\begin{eqnarray*}
\frac{\partial^2 l_W(\eta_0)}{\partial \gamma^2}=0, \ \frac{\partial^2 l_W(\eta_0)}{\partial \gamma \partial \alpha^\ast}=0, \ \frac{\partial^2 l_W(\eta_0)}{\partial \gamma \partial \beta^\top}=0, \  \frac{\partial^2 l_W(\eta_0)}{\partial \gamma \partial \mu^\top}= 0,
\end{eqnarray*}
\begin{eqnarray*}
\frac{\partial^2 l_W(\eta_0)}{\partial \gamma \partial \nu^\top}= -(1+\rho)\sum_{i=1}^n \frac{\partial H(x_i ; \theta_0)/\partial \gamma}{\Delta_i}, \ \frac{\partial^2 l_W(\eta_0)}{\partial {\alpha^\ast}^2}= -\rho \sum_{i=1}^n \frac{\delta_i}{\Delta^2_i},
\end{eqnarray*}
\begin{eqnarray*}
\frac{\partial^2 l_W(\eta_0)}{\partial \alpha^\ast \partial \beta^\top}= - \rho \sum_{i=1}^n \frac{\delta_i x_i}{\Delta^2_i}, \ 
\frac{\partial^2 l_W(\eta_0)}{\partial \alpha^\ast \partial \mu^\top} = 0,
\end{eqnarray*}
\begin{eqnarray*}
\frac{\partial^2 l_W(\eta_0)}{\partial \alpha^\ast \partial \nu^\top}= -(1+\rho)\sum_{i=1}^n \frac{\partial H(x_i ; \theta_0)/\partial \alpha^\ast}{\Delta_i} + \rho(1+\rho)\sum_{i=1}^n \frac{\delta_i H(x_i,\theta_0)}{\Delta^2_i},
\end{eqnarray*}
\begin{eqnarray*}
\frac{\partial^2 l_W(\eta_0)}{\partial \beta \partial \beta^\top}= -\rho \sum_{i=1}^n \frac{\delta_i x_i {x_i}^\top}{\Delta^2_i}, \ 
\frac{\partial^2 l_W(\eta_0)}{\partial \beta \partial \mu^\top} = 0,
\end{eqnarray*}
\begin{eqnarray*}
\frac{\partial^2 l_W(\eta_0)}{\partial \beta \partial \nu^\top} = -(1+\rho)\sum_{i=1}^n \frac{\partial H(x_i ; \theta_0)/\partial \beta}{\Delta_i} + \rho(1+\rho)\sum_{i=1}^n \frac{\delta_i x_i H^\top(x_i,\theta_0)}{\Delta^2_i}
\end{eqnarray*}
\begin{eqnarray*}
\frac{\partial^2 l_W(\eta_0)}{\partial \mu \partial \mu^\top}=-NW^{-1}, \ \frac{\partial^2 l_W(\eta_0)}{\partial \mu \partial \nu^\top}= -(1+\rho)\sum_{i=1}^n \frac{\partial H(x_i ; \theta_0)/\partial \mu}{\Delta_i},
\end{eqnarray*}
\begin{eqnarray*}
\frac{\partial^2 l_W(\eta_0)}{\partial \nu \partial \nu^\top}= (1+\rho)^2 \sum_{i=1}^n \frac{H(x_i ; \theta_0) H^\top (x_i ; \theta_0)}{\Delta^2_i}.
\end{eqnarray*}

Applying the first-order Taylor expansion on $\partial l_W(\hat{\eta}_W)/ \partial \eta$ and using the law of large numbers and the consistency results from Theorem \ref{thm1}, we obtain that
\begin{eqnarray*}
0=\frac{\partial l_W(\hat{\eta}_W)}{\partial \eta}=\frac{\partial l_W(\eta_0)}{\partial \eta}+\mathsf{E}\left\{\frac{\partial^2 l_W(\eta_0)}{\partial \eta \partial \eta^{\top}}\right\}\left(\hat{\eta}_W-\eta_0\right)+o_p\left(n^{1 / 2}\right).
\end{eqnarray*}
Then
\begin{eqnarray}\label{asym}
\hat{\eta}_W-\eta_0=-\left[\mathsf{E}\left\{\frac{\partial^2 l_W\left(\eta_0\right)}{\partial \eta \partial \eta^{\top}}\right\}\right]^{-1} \frac{\partial l_W\left(\eta_0\right)}{\partial \eta}+o_p\left(n^{-1 / 2}\right)
\end{eqnarray}
where 
\begin{equation*} 
\frac{1}{n} \mathsf{E}\left\{\frac{\partial^2 l\left(\theta_0, \nu_0\right)}{\partial \eta \partial \eta^{\top}}\right\}=\left(\begin{array}{ccccc}
0 & 0 & 0 & 0 & -A_{15} \\
0 & -A_{22} & -A_{23} & 0 & -A_{25} \\
0 & -A_{32} & -A_{33} & 0 & -A_{35} \\
0 & 0 & 0 & -A_{44} & -A_{45} \\
-A_{51} & -A_{52} & -A_{53} & -A_{54} & A_{55}
\end{array}\right) 
\end{equation*}
with
\begin{eqnarray*}
A_{15}= A^{\top}_{51} = \mathsf{E}_0\left\{\frac{\partial H^{\top}(X;\theta_0)}{\partial \gamma}\right\}, \quad
A_{22} = \frac{\rho}{1+\rho} \mathsf{E}_0\left(\frac{\delta_0}{\Delta_0}\right), \quad
A_{23}=A_{32}^{\top}=\frac{\rho}{1+\rho} \mathsf{E}_0\left(\frac{\delta_0 X^{\top}}{\Delta_0}\right),
\end{eqnarray*}
\begin{eqnarray*}
A_{25}=A_{52}^{\top}=\mathsf{E}_0\left\{\frac{\partial H^{\top}\left(X ; \theta_0\right)}{\partial \alpha^\ast}\right\}+\mathsf{E}_0\left\{\frac{H^{\top}\left(X ; \theta_0\right)}{\Delta_0}\right\}, \quad
A_{33}=\frac{\rho}{1+\rho}\mathsf{E}_0\left(\frac{\delta_0 X X^{\top}}{\Delta_0}\right),
\end{eqnarray*}
\begin{eqnarray*}
A_{35}=A_{53}^{\top}=\mathsf{E}_0\left\{\frac{\partial H^{\top}\left(X ; \theta_0\right)}{\partial \beta}\right\}-\rho E_0\left\{\frac{\delta_0 X H^{\top}\left(X ; \theta_0\right)}{\Delta_0}\right\},\quad
A_{44} = \frac{N}{n}W^{-1},
\end{eqnarray*}
\begin{eqnarray*}
A_{45} = A^{\top}_{54} = \mathsf{E}_0\left\{\frac{\partial H^{\top}(X;\theta_0)}{\partial \mu}\right\},\quad
A_{55} =(1+\rho) \mathsf{E}_0\left\{\frac{H\left(X ; \theta_0\right) H^{\top}\left(X ; \theta_0\right)}{\Delta_0}\right\}, 
\end{eqnarray*}
where $\mathsf{E}_0$ means taking expectation with respect to the conditional density $f_0(x)$. Here we omit the details on how to get $A_{ij}$, which is similar to the proof of Theorem 1 in Qin (2014), mainly using Lemma 1 in that paper to integrate the second order derivation on $Y=0$ and $Y=1$ separately.

To get the asymptotic form of (\ref{asym}), by doing some calculation, we have that
\begin{eqnarray}\label{af1}
    \left(\begin{array}{cccc}
0 & 0 & 0 & 0  \\
0 & -A_{22} & -A_{23} & 0  \\
0 & -A_{32} & -A_{33} & 0 \\
0 & 0 & 0 & -A_{44}  \\
\end{array}\right) (\hat{\theta}_W - \theta_0) -  \left(\begin{array}{c}
A_{15} \\
A_{25} \\
A_{35} \\
A_{45} \\
\end{array}\right)(\hat{\nu}_W - \nu_0) =-n^{-1}\left(\begin{array}{c}
0 \\
S_{2n} \\
S_{3n} \\
S_{4n} 
\end{array}\right) + o_p(n^{-1/2})
\end{eqnarray}
and
\begin{eqnarray}\label{af2}
-(A_{51},A_{52},A_{53},A_{54})(\hat{\theta}_W - \theta_0) + A_{55}(\hat{\nu}_W - \nu_0) = -n^{-1}S_{5n} + o_p(n^{-1/2}),
\end{eqnarray}
where $$S_{2n} =\frac{\partial l_W(\eta_0)}{\partial \alpha^*}= n_1 - \rho \sum_{i=1}^n \frac{\delta_i}{\Delta_i}, \ S_{3n} = \frac{\partial l_W(\eta_0)}{\partial \beta} = \sum_{i=1}^n y_i x_i - \rho \sum_{i=1}^n \frac{\delta_i x_i}{\Delta_i},$$ $$S_{4n} = \frac{\partial l_W(\eta_0)}{\partial \mu} = NW^{-1}(\tilde{\mu}-\mu_0),$$ and $$S_{5n} =\frac{\partial l_W(\eta_0)}{\partial \nu}=-(1+\rho)\sum_{i=1}^n \frac{H(x_i ; \theta_0)}{\Delta_i}.$$ 
Hence, from (\ref{af2}) we have that
\begin{equation*}
\hat{\nu}_W - \nu_0 = A^{-1}_{55}(A_{51},A_{52},A_{53},A_{54})(\hat{\theta}_W - \theta_0) - A^{-1}_{55}(n^{-1}S_{n5}) + o_p(n^{-1/2}).
\end{equation*}
Replacing it into (\ref{af1}), we have that
\begin{eqnarray}\label{aysm2}
& &\left(\begin{array}{cccc}
A_{15}A^{-1}_{55}A_{51} & A_{15}A^{-1}_{55}A_{52} & A_{15}A^{-1}_{55}A_{53} & A_{15}A^{-1}_{55}A_{54}  \\
A_{25}A^{-1}_{55}A_{51} & A_{25}A^{-1}_{55}A_{52}+A_{22} & A_{25}A^{-1}_{55}A_{53}+A_{23} & A_{25}A^{-1}_{55}A_{54}  \\
A_{35}A^{-1}_{55}A_{51} & A_{35}A^{-1}_{55}A_{52}+A_{32} & A_{35}A^{-1}_{55}A_{53}A_{33} & A_{35}A^{-1}_{55}A_{54} \\
A_{45}A^{-1}_{55}A_{51} & A_{45}A^{-1}_{55}A_{52} & A_{45}A^{-1}_{55}A_{53} & A_{45}A^{-1}_{55}A_{54}+A_{44}  \\
\end{array}\right) (\hat{\theta}_W - \theta_0)\nonumber\\
&=&\left(
\begin{array}{cccc}
    0 & 0 & 0 &  A_{15}A^{-1}_{55} \\
    1 & 0 & 0 &  A_{25}A^{-1}_{55} \\
    0 & I & 0 &  A_{35}A^{-1}_{55} \\
    0 & 0 & I &  A_{45}A^{-1}_{55} 
\end{array} 
\right)(n^{-1}S_{n}) + o_p(n^{-1/2}),
\end{eqnarray}
where $S_n=(S_{2n}, S_{3n}^\top, S_{4n}^\top, S_{5n}^\top)^\top$. Define
\begin{eqnarray*}
U_W = \left(\begin{array}{cccc}
0 & 0 & 0 & A_{15} \\
A_{22} & A_{23} & 0 & A_{25} \\
A_{32} & A_{33} & 0 & A_{35} \\
0 & 0 & A_{44} & A_{45}
\end{array}\right), \  M_W =  \left(\begin{array}{cccc}
A_{22} & A_{23} & 0 & 0 \\
A_{32} & A_{33} & 0 & 0 \\
0 & 0 & A_{44} & 0 \\
0 & 0 & 0 & A_{55}
\end{array}\right),
\end{eqnarray*}
and $J_W=U_W M_W^{-1} U_W^{\top}$. Then by some algebra, it can be verified that (\ref{aysm2}) can be written as 
\begin{eqnarray*}
J_W(\hat{\theta}_W-\theta_0)=U_W M_W^{-1}(n^{-1} S_n)+o_p(n^{-1/2}).
\end{eqnarray*}
Therefore,
\begin{eqnarray}\label{asym3}
\sqrt{n}(\hat{\theta}_W-\theta_0)=J_W^{-1} U_W M_W^{-1}(n^{-1 / 2} S_n)+o_p(1).
\end{eqnarray}
Hence, we conclude that $\sqrt{n}(\hat{\theta}_W-\theta_0)\to N(0, \Sigma_W)$ in distribution, where
\begin{eqnarray}\label{sigmaW}
\Sigma_W=\{J_W^{-1} U_W M_W^{-1}\}\left\{\mathsf{Var}(n^{-1 / 2} S_n)\right\}\{M_W^{-1} U_W^{\top} J_W^{-1}\}
\end{eqnarray}
The last step is to obtain the specific form of $\mathsf{Var}(n^{-1 / 2} S_n)$. Note that except for $S_{4n}$, the forms of $S_{2n}$, $S_{3n}$, and $S_{5n}$ are similar those given in Lemma 3 of Qin(2014). From {\it Condition 4}, we have  that$\sqrt{N}(\tilde{\mu}-\mu_0) \sim N(0,V)$. Thus
\begin{eqnarray*}
\mathsf{Var}(S_{4n}) = \mathsf{Var}(NW^{-1}(\tilde{\mu}-\mu_0)) = NW^{-1} V W^{-1}
\end{eqnarray*}
and 
\begin{eqnarray*}
\mathsf{Cov}(S_{4n},S_{jn}) = 0 \quad \text{for } j=2,3,5.
\end{eqnarray*}
Therefore,
\begin{eqnarray*}
\frac{1}{n} 
\mathsf{Var}\left(S_n\right)=\left(\begin{array}{cccc}
A_{22} & A_{23} & 0 & 0\\
A_{32} & A_{33} & 0 & 0\\
0 & 0 & \frac{N}{n}W^{-1} V W^{-1} & 0 \\
0 & 0 &0 & A_{55}
\end{array}\right)-\frac{(1+\rho)^2}{\rho}\left(\begin{array}{c}
A_{22} \\
A_{32} \\
0\\
A_{52}+A_{51}
\end{array}\right)\left(\begin{array}{c}
A_{22} \\
A_{32} \\
0\\
A_{52}+A_{51}
\end{array}\right)^{\top} .
\end{eqnarray*}
Set 
\begin{eqnarray*}
c = \left(\begin{array}{c}
A_{22} \\
A_{32} \\
0\\
A_{52}+A_{51}
\end{array}\right)
\end{eqnarray*} 
and
\begin{eqnarray*}
D_W = 
\left(\begin{array}{cccc}
0 & 0 & 0 & 0\\
0 & 0 & 0 & 0\\
0 & 0 & -\frac{N}{n}W^{-1}+\frac{N}{n}W^{-1} V W^{-1} & 0 \\
0 & 0 &0 & 0
\end{array}\right)
\end{eqnarray*} 
Finally, we obtain that
\begin{eqnarray*}
\mathsf{Var}\left(n^{-1/2}S_n\right) = M_W + D_W  - \frac{(1+\rho)^2}{\rho} c c^{\top}.
\end{eqnarray*} 
Replace this into (\ref{sigmaW}), we can obtain the final form of $\Sigma_W$. This completes the proof of Theorem 2.\qed

\medskip

\noindent{\it Proof of Corollary \ref{cor1}:} \ For convenience, we define 
$S_{4n,W} = NW^{-1}(\tilde{\mu}-\mu_0)$, $S_{4n,V} = NV^{-1}(\tilde{\mu}-\mu_0)$, and $C = ((1+\rho)^2/\rho) c c^\top$. When $W = V$, we have that $\partial^2 l_{V}(\eta_0)/\partial \mu \partial \mu^\top=-NV^{-1}$, and we write $A_{44}$ as $A_{44,V} = (N/n) V^{-1}$ and $\mathsf{Var}(n^{-1/2}S_{4n,V}) =(N/n)V^{-1} = A_{44,V}$. We also denote the corresponding $J$, $U$, and $M$ as $J_{V}$, $U_{V}$, and $M_{V}$. When $W \neq V$, we have that $\partial^2 l_{W}(\eta_0)/\partial \mu \partial \mu^\top=-NW^{-1}$ and we write $A_{44}$ as $A_{44,W} = (N/n) W^{-1}$ and $\mathsf{Var}(n^{-1/2}S_{4n,W}) =(N/n) W^{-1} V W^{-1}  \neq A_{44,W}$.

For the above notation, we have $U_W \rightarrow U_V $, $M_W \rightarrow M_V $ and $J_W \rightarrow J_V $ as $n$ goes to infinity. Meanwhile, from Theorem 2, the asymptotic variance-covariance matrix of $\sqrt{n}(\hat{\theta}_V-\theta_0)$ equals to $J_V^{-1}U_V M_V^{-1}(M_V - C) M_V^{-1} U_V^{\top} J_V^{-1}$. Let $S_{n,V}$ and $S_{n,W}$ denote $S_n$ with $S_{4n,V}$ and $S_{4n,W}$, respectively. From 
\begin{eqnarray*}
\sqrt{n}(\hat{\theta}_V - \theta_0) = J_V^{-1}U_V M_V^{-1}(n^{-1/2} S_{n,V}) + o_p(1),
\end{eqnarray*}
we have that
\begin{eqnarray*}
\sqrt{n}\hat{\theta}_V &=& \sqrt{n}\theta_0 + J_V^{-1}U_V M_V^{-1}(n^{-1/2}S_{n,V}) + o_p(1) \\
&=& n^{1/2}\theta_0 + J_V^{-1}U_V M_V^{-1}\left(n^{-1/2}S_{n,W} + \left(\begin{array}{c}
0 \\
0 \\
n^{-1 / 2} N\left(W^{-1}-V^{-1}\right)\left(\mu_0-\tilde{\mu}\right) \\
0
\end{array}\right)\right) + o_p(1).
\end{eqnarray*}
Hence, it can be argued that
\begin{eqnarray*}
& & \mathsf{Cov}(\sqrt{n}\hat{\theta}_W, \sqrt{n}\hat{\theta}_V) \\
& \rightarrow& J_W^{-1}U_W M_W^{-1} \left(\mathsf{Var}(n^{-1/2}S_{n,W}) + \left(\begin{array}{cccc}
0 & 0 & 0 & 0 \\
0 & 0 & 0 & 0 \\
0 & 0  & -\frac{N}{n} W^{-1} V \left(W^{-1}-V^{-1}\right)& 0\\
0 & 0 & 0 & 0 
\end{array}\right)\right) M_V^{-1} U_V^{\top} J_V^{-1}  \\
&=& J_W^{-1}U_W M_W^{-1}(M_W - C) M_V^{-1} U_V^{\top} J_V^{-1} \\
&\rightarrow& J_V^{-1}U_V M_V^{-1}(M_V - C) M_V^{-1} U_V^{\top} J_V^{-1}
\end{eqnarray*}
which is just the asymptotic variance-covariance matrix of $\sqrt{n}(\hat{\theta}_V-\theta_0)$. Moreover
\begin{eqnarray*}
\mathsf{Cov}(\sqrt{n}\hat{\theta}_W - \sqrt{n}\hat{\theta}_V) &=&\mathsf{Cov}(\sqrt{n}\hat{\theta}_W) + \mathsf{Cov}(\sqrt{n}\hat{\theta}_V) -2 \mathsf{Cov}(\sqrt{n}\hat{\theta}_W, \sqrt{n}\hat{\theta}_V) \\
& \rightarrow&\mathsf{Cov}(\sqrt{n}\hat{\theta}_W) + \mathsf{Cov}(\sqrt{n}\hat{\theta}_V) -2\mathsf{Cov}(\sqrt{n}\hat{\theta}_V) \\
& =&\mathsf{Cov}(\sqrt{n}\hat{\theta}_W) -\mathsf{Cov}(\sqrt{n}\hat{\theta}_V) \\
&\geqslant& 0.
\end{eqnarray*}
Hence we get
\begin{eqnarray*}
    \mathsf{Cov}(\sqrt{n}\hat{\theta}_W) \geqslant \mathsf{Cov}(\sqrt{n}\hat{\theta}_V) 
\end{eqnarray*}
Therefore, we can see that the variance-covariance matrix of the limiting distribution of $\sqrt{n}(\hat{\theta}_W-\theta_0)$ attains the minimum at $W=V$. Meanwhile, when $W=V$, we can simplify (\ref{sigmaW}) as
\begin{eqnarray}\label{sigmaV}
\Sigma_V &=&\{J_V^{-1} U_V M_V^{-1}\}\left\{Var(n^{-1 / 2} S_{n,V})\right\}\{M_V^{-1} U_V^{\top} J_V^{-1}\} \notag \\
&=&J_V^{-1} - \frac{(1+\rho)^2}{\rho}(1,1,0,0)^\top (1,1,0,0).
\end{eqnarray}

Finally, when $W$ takes the form of any consistent estimator $\hat{V}$, we have that$U_{\hat{V}} \rightarrow U_V $, $M_{\hat{V}} \rightarrow M_V $, and $J_{\hat{V}} \rightarrow J_V $ in probability, respectively. Hence, we can show that $\mathsf{Cov}(\sqrt{n}\hat{\theta}_W, \sqrt{n}\hat{\theta}_{\hat{V}}) \rightarrow J_W^{-1}U_W M_W^{-1}(M_W - C) M_{\hat{V}}^{-1} U_{\hat{V}}^{\top} J_{\hat{V}}^{-1} \rightarrow \mathsf{Cov}(\sqrt{n}\hat{\theta}_{\hat{V}}) $. Similarly, we can obtain that $\mathsf{Cov}(\sqrt{n}\hat{\theta}_W - \sqrt{n}\hat{\theta}_{\hat{V}}) \geqslant 0$ and the variance-covariance matrix of the limiting distribution of $\sqrt{n}(\hat{\theta}_{\hat{V}}-\theta_0)$ remains to be $\Sigma_V$.\qed

\medskip

\noindent{\it Proof of Corollary \ref{cor2}:} \ $H(x;\theta)$ is a $q+1$ dimension vector function of $x$. For $2 < r \leqslant q+1$, let $U_r$, $M_r$, $J_r$, and $\Sigma_r$ denote the matrices corresponding to $U_{\hat{V}}$, $M_{\hat{V}}$, $J_{\hat{V}}$, and $\Sigma_{\hat{V}}$, only using the first $r$ estimating equations of $H(x;\theta)$, respectively. Because $\hat{V} $ is a consistent estimator of $V$, which means $\hat{V}^{-1}V\hat{V}^{-1} \rightarrow \hat{V}^{-1}$. Then we have $D_{\hat{V}} \rightarrow 0$ and 
$\Sigma_{\hat{V}} = J_{\hat{V}}^{-1} - \frac{(1+\rho)^2}{\rho}(1,1,0,0)^\top (1,1,0,0)$

In order to demonstrate the asymptotic variance-covariance matrix cannot decrease if any function in $h(x)$ is dropped, it suffices to show that 
\begin{eqnarray*}
\Sigma_r \leqslant \Sigma_{r-1}
\end{eqnarray*}
or equivalently
\begin{eqnarray*}
J_r \geqslant J_{r-1}.
\end{eqnarray*}
By the proof of Corollary of  Qin and Lawless (1994;), we can get that
\begin{eqnarray*}
M_r^{-1} \geqslant\left(\begin{array}{cc}
M_{r-1}^{-1} & 0 \\
0 & 0
\end{array}\right)
\end{eqnarray*}
Write $U_r=\left(U_{r-1}, u_r\right)$. Therefore, it can be seen that
\begin{eqnarray*}
J_r=U_r M_r^{-1} U_r^\top \geqslant\left(U_{r-1}, u_r\right)\left(\begin{array}{cc}
M_{r-1}^{-1} & 0 \\
0 & 0
\end{array}\right)\left(U_{r-1}, u_r\right)^\top =U_{r-1} M_{r-1}^{-1} U_{r-1}^\top =J_{r-1}.
\end{eqnarray*}
By the fact $\Sigma_{\hat{V}} = J_{\hat{V}}^{-1} - \frac{(1+\rho)^2}{\rho}(1,1,0,0)^\top (1,1,0,0)$, the conclusion of Corollary \ref{cor2} can be obtained.\qed

\medskip

\noindent{\it The specific form of $\Sigma_I$}: We make some minor adjustments to the notation of the parameters. We set $\hat{\theta}_I=(\hat{\gamma}, \hat{\alpha}^\ast, \hat{\beta}^\top)^\top$ and the true value $\theta_{I,0}=(\gamma_0,\alpha^\ast_0,\beta_0^\top)^\top$. Similar to the notation of $U_W$, $M_W$, and $J_W$, define
\begin{eqnarray*}
U_I = \left(\begin{array}{ccc}
0 & 0 & A_{15} \\
A_{22} & A_{23} & A_{25} \\
A_{32} & A_{33} & A_{35} 
\end{array}\right), \  M_I =  \left(\begin{array}{ccc}
A_{22} & A_{23} & 0  \\
A_{32} & A_{33} & 0  \\
0 & 0  & A_{55}
\end{array}\right),
\end{eqnarray*}
and $J_I=U_I M_I^{-1} U_I^{\top}$, where $H(x,\theta)$ is also adjusted corresponding to $\theta_I$. Then, by some algebra similar to the derivation of (\ref{asym3}), we have that
\begin{eqnarray*}
\sqrt{n}(\hat{\theta}_I-\theta_0)=J_I^{-1} U_I M_I^{-1}(n^{-1 / 2} S_n)+o_p(1).
\end{eqnarray*}
Hence, we can conclude that $\sqrt{n}(\hat{\theta}_I-\theta_0)\to N(0, \Sigma_I)$ in distribution, where $\Sigma_I$ can be simplified as follows:
\begin{eqnarray*}
\Sigma_I=J_I^{-1} - \frac{(1+\rho)^2}{\rho}(1,1,0,0)^\top (1,1,0,0).
\end{eqnarray*}

\subsection{Estimation of the asymptotic variance-covariance matrices}\label{a2}

Here we use plug-in method to give out the specific form of the estimator of $\Sigma_W$, denoted by $\hat{\Sigma}_W$. According to White (1982) we replace $\theta_0$ with its consistent estimates $\hat{\theta}_W$ in $H(x;\theta_0)$ and its derivation $\partial H(x;\theta_0)/\partial \theta$. For $A_{ij},i,j = 1,2,...,5$, we use the sample mean for the control sample to estimate $\mathsf{E}_0$, and use the notation $x_{0i}$'s to represent those $x_i$'s with $y_i=0$, $i = 1,2,...,n_0$. Then the estimators of $A_{ij}$ are given as follows:
\begin{eqnarray*}
\hat{A}_{15}=\hat{A}^{\top}_{51} = \frac{1}{n_0}\sum_{i=1}^{n_0}\left(\frac{\partial H^{\top}(x_{0i};\hat{\theta}_W)}{\partial \gamma}\right), \ \hat{A}_{22}=\frac{\rho}{1+\rho} \frac{1}{n_0}\sum_{i=1}^{n_0}\left(\frac{\delta_i}{\Delta_i}\right),
\end{eqnarray*}
\begin{eqnarray*}
\hat{A}_{23}=\hat{A}_{32}^{\top}=\frac{\rho}{1+\rho} \frac{1}{n_0}\sum_{i=1}^{n_0}\left(\frac{\delta_i x_{0i}^{\top}}{\Delta_i}\right),
\end{eqnarray*}
\begin{eqnarray*}
\hat{A}_{25}=\hat{A}_{52}^{\top}=\frac{1}{n_0}\sum_{i=1}^{n_0}\left(\frac{\partial H^{\top}\left(x_{0i} ; \hat{\theta}_W\right)}{\partial \alpha^\ast}\right)+\frac{1}{n_0}\sum_{i=1}^{n_0}\left(\frac{H^{\top}\left(x_{0i} ; \hat{\theta}_W\right)}{\Delta_i}\right), 
\end{eqnarray*}
\begin{eqnarray*}
\hat{A}_{33}=\frac{\rho}{1+\rho}\frac{1}{n_0}\sum_{i=1}^{n_0}\left(\frac{\delta_i x_{0i} x_{0i}^{\top}}{\Delta_i}\right),
\end{eqnarray*}
\begin{eqnarray*}
\hat{A}_{35}=\hat{A}_{53}^{\top}=\frac{1}{n_0}\sum_{i=1}^{n_0}\left(\frac{\partial H^{\top}\left(x_{0i} ; \hat{\theta}_W\right)}{\partial \beta}\right)-\rho \frac{1}{n_0}\sum_{i=1}^{n_0}\left(\frac{\delta_i x_{0i} H^{\top}\left(x_{0i} ; \hat{\theta}_W\right)}{\Delta_i}\right),
\end{eqnarray*}
\begin{eqnarray*}
\hat{A}_{44} = \frac{N}{n}W^{-1}, \quad \hat{A}_{45} = \hat{A}^{\top}_{54} = \frac{1}{n_0}\sum_{i=1}^{n_0}\left(\frac{\partial H^{\top}(x_{0i};\hat{\theta}_W)}{\partial \mu}\right),
\end{eqnarray*}
\begin{eqnarray*}
\hat{A}_{55} =(1+\rho) \frac{1}{n_0}\sum_{i=1}^{n_0}\left(\frac{H\left(x_{0i} ; \hat{\theta}_W\right) H^{\top}\left(x_{0i} ; \hat{\theta}_W\right)}{\Delta_i}\right).
\end{eqnarray*}
By replacing all the $A_{ij}$'s by the corresponding $\hat{A}_{ij}$'s, we denote the estimator of $U_W$, $M_W$, $J_W$, and ${c}$ by $\hat{U}_W$, $\hat{M}_W$, $\hat{J}_W$, and $\hat{c}$, respectively. For estimating $D_W$, we use $\hat{V}$ to estimate $V$, and the corresponding estimator is denoted by $\hat{D}_W$. Then a plug-in estimator of $\Sigma_W$ in (\ref{sigmaW}) is given by 
\begin{eqnarray*}
\hat{\Sigma}_W=\left(\hat{J}_W^{-1} \hat{U}_W \hat{M}_W^{-1}\right)\left(\hat{M}_W+\hat{D}_W-\frac{(1+\rho)^2}{\rho}\hat{c}\hat{c}^\top\right)\left(\hat{M}_W^{-1} \hat{U}_W^{\top} \hat{J}_W^{-1}\right).
\end{eqnarray*}

\bigskip

\noindent {\bf Acknowledgment}

Ming Zheng's research is supported by the National Natural Science Foundation of China Grants (12271106). Wen Yu's research is supported by the National Natural Science Foundation of China Grants (12071088).

\end{document}